\documentclass[
aps,
prd,
tightenlines,
superscriptaddress,
nofootinbib,
floatfix,
groupedaddress,
preprintnumbers,
longbibliography,
nofootinbib,
amsmath,
amsfonts,
amssymb]
{revtex4}
\usepackage{notoccite}
\usepackage{graphicx,
longtable,
color
}

\usepackage[dvipsnames]{xcolor}
\usepackage[colorlinks=true,breaklinks=true]{hyperref}
\usepackage[utf8]{inputenc}
\hypersetup{
    colorlinks=true,
    citecolor = green,
    linkcolor=red,
    filecolor=magenta,      
    urlcolor=blue,
}
\hypersetup{allcolors=[rgb]{0.0 0.0 0.9},linkcolor=[rgb]{0.75 0.05 0.05}}

\usepackage{url}

\usepackage{xspace}
\usepackage[normalem]{ulem}
\newcommand{\nua}[1]{\ensuremath{\rlap{\kern-1.0pt\ensuremath{\overset{\scriptscriptstyle(-)}{\phantom{\nu}}}}{\ensuremath{{\nu}_{#1}}}}\xspace}

\begin{document}

\preprint{IPPP/20/16}

\title{Interpretation of  NO$\nu$A and T2K data in the presence of a light sterile neutrino}%

\author{		Sabya Sachi Chatterjee}
\email{sabya.s.chatterjee@durham.ac.uk}
\affiliation{		Institute for Particle Physics Phenomenology, Department of Physics, Durham University, Durham, DH1 3LE, UK}

\author{		Antonio Palazzo}
\email{antonio.palazzo@ba.infn.it}
\affiliation{ 	Dipartimento Interateneo di Fisica ``Michelangelo Merlin,'' Via Amendola 173, 70126 Bari, Italy}
\affiliation{ 	Istituto Nazionale di Fisica Nucleare, Sezione di Bari, Via Orabona 4, 70126 Bari, Italy}


\begin{abstract}
We study in detail the impact of a light sterile neutrino in the interpretation of the latest  data of the long baseline
experiments NO$\nu$A and T2K, assessing the robustness/fragility of the estimates of the standard 3-flavor parameters
with respect to the perturbations induced in the 3+1 scheme. We find that all the basic features of the 3-flavor analysis, including the weak 
indication ($\sim$1.4$\sigma$) in favor of the inverted neutrino mass ordering, the preference for values of the CP-phase $\delta_{13} \sim 1.2\pi$,
and the substantial degeneracy of the two octants of $\theta_{23}$, all remain basically unaltered in the 4-flavor scheme. Our analysis also demonstrates that it is possible to attain some constraints on the new CP-phase $\delta_{14}$. Finally, we point out that, differently from
non-standard neutrino interactions, light sterile neutrinos are not capable to alleviate the tension recently emerged between NO$\nu$A and T2K 
in the appearance channel.

\end{abstract}
\maketitle

\section{Introduction}

Neutrino mass and mixing  have been firmly established by several
experiments using natural and artificial neutrino sources.
The 3-flavor framework is recognized as the sole scheme able 
to describe all the measurements obtained at baselines longer than $\sim 100$ meters.
Despite its huge success, however, the standard 3-flavor scheme might not constitute the ultimate
 description of neutrino properties.
 As a matter of fact, several anomalies have been found in short baseline experiments (SBL), which cannot
be described in the 3-flavor scenario (for reviews on the subject see~\cite{Abazajian:2012ys,Palazzo:2013me,Gariazzo:2015rra,Giunti:2015wnd,Giunti:2019aiy,Boser:2019rta,Capozzi:2016vac,Gariazzo:2017fdh,Dentler:2018sju,Diaz:2019fwt}).
The hints derive from the accelerator experiments LSND~\cite{Aguilar:2001ty} and MiniBooNE~\cite{Aguilar-Arevalo:2018gpe}, and from the so-called reactor~\cite{Mention:2011rk} and Gallium~\cite{Hampel:1997fc,Abdurashitov:2005tb} anomalies. More recently,
the nuclear reactor data from NEOS~\cite{Ko:2016owz}, DANSS~\cite{Danilov:2019aef} and
Neutrino-4~\cite{Serebrov:2018vdw}, have provided indications in the same direction.
Limits on light sterile neutrinos have been
also found exploiting solar neutrinos~\cite{Palazzo:2011rj,Palazzo:2012yf,Giunti:2009xz},
the long-baseline (LBL) experiments MINOS, MINOS+~\cite{MINOS:2016viw,Adamson:2017uda}, 
NO$\nu$A~\cite{Adamson:2017zcg} and T2K~\cite{Abe:2019fyx}, the reactor experiments Daya Bay~\cite{An:2016luf},
PROSPECT~\cite{Ashenfelter:2018iov} and STEREO~\cite{AlmazanMolina:2019qul}, and also  the atmospheric data collected
at Super-Kamiokande~\cite{Abe:2014gda}, IceCube~\cite{TheIceCube:2016oqi,Aartsen:2017bap} and ANTARES~\cite{Albert:2018mnz}.

In the 3-flavor framework three mass eigenstates $\nu_i$ with masses
$m_i\, (i = 1,2,3)$ are introduced, together with three mixing angles ($\theta_{12},\theta_{23}, \theta_{13}$), and one CP-phase $\delta_{13}$.
The neutrino mass ordering (NMO) is dubbed  normal (NO)  if $m_3>m_{1,2}$ or inverted (IO) if  $m_3<m_{1,2}$.
The two mass-squared splittings $\Delta m^2_{21} \equiv m^2_2 - m^2_1$ and $\Delta m^2_{31} \equiv m^2_3 - m^2_1$
are too small to give rise to detectable effects in SBL setups. Therefore, new neutrino species, having 
much bigger mass-squared differences $O(\mathrm{eV}^2)$ must be introduced to explain the SBL anomalies. The new 
hypothetical neutrino states are assumed to be sterile, i.e. singlets of the standard model gauge group. 
Several new and more sensitive SBL experiments are underway to put under test such an intriguing hypothesis 
(see the review in~\cite{Boser:2019rta}), which, if confirmed would constitute a tangible evidence of physics 
beyond the standard model.
 In the minimal extension of the 3-flavor framework, the so-called  $3+ 1$ scheme,
only one sterile species is introduced.  In this scenario, one supposes that 
one mass eigenstate $\nu_4$ exists, weakly mixed with the active neutrino flavors ($\nu_e, \nu_\mu, \nu_\tau$) and 
separated from the standard mass eigenstates $(\nu_1, \nu_2, \nu_3)$ by a $O(\mathrm{eV}^2)$ difference.
The 3+1 framework is governed by six mixing angles and three (Dirac) CP-violating phases. Therefore, in the eventuality of 
a discovery of a light sterile species, we would face the challenging task of identifying six new properties
[3 mixing angles ($\theta_{14}$, $\theta_{24}$, $\theta_{34}$), 2 CP-phases ($\delta_{14}$, $\delta_{34}$), and
the mass-squared splitting $\Delta m^2_{41}\equiv m^2_4 - m^2_1$].%

As first pointed out in Ref.~\cite{Klop:2014ima}, in the presence of a light sterile neutrino, 
the $\nu_\mu \to \nu_e$ conversion probability probed by the long baseline (LBL) facilities
entails a new term engendered by the interference among the atmospheric frequency
and the new frequency connected to the sterile species. The 
oscillations induced by the new frequency are very fast and are completely smeared out by the finite 
energy resolution of the detector. Notwithstanding, the fourth neutrino leaves observable traces 
in the conversion probability.
This makes the LBL experiments able to probe the new CP-phases entailed by the 3+1 scheme. 
The recent 4-flavor analyses of NO$\nu$A and T2K data, have already pointed out
that such two experiments have some sensitivity to one of such CP-phases~\cite{Klop:2014ima,Palazzo:2015gja,Capozzi:2016vac}.
Likewise, the sensitivity study carried out in~\cite{Agarwalla:2016mrc} has pointed out that the 
discovery potential of the CP-phases is expected to increase when NO$\nu$A and 
T2K will attain their final exposures, and will be further improved in the new-generation LBL
facilities DUNE~\cite{Hollander:2014iha,Berryman:2015nua,Gandhi:2015xza,Agarwalla:2016xxa,Agarwalla:2016xlg,Coloma:2017ptb,Choubey:2017cba}, T2HK~\cite{Choubey:2017cba,Choubey:2017ppj,Agarwalla:2018nlx},
T2HKK~\cite{Choubey:2017cba,Haba:2018klh},
and ESS$\nu$SB~\cite{Agarwalla:2019zex}
(for a recent review on the topic see Ref.~\cite{Palazzo:2020tye}).
In this work, we analyze the latest data provided by NO$\nu$A and T2K, 
and we assess the robustness of the 3-flavor estimates of the oscillation parameters 
in the enlarged 3+1 scheme. In addition, we analyze the indications on the
new CP-phase $\delta_{14}$. Further studies on the the impact of sterile neutrinos
in LBL experiments, mostly focused on future experiments, can be found in~\cite{Donini:2001xy,Donini:2001xp,Donini:2007yf,Dighe:2007uf,Donini:2008wz,Yasuda:2010rj,Meloni:2010zr,Bhattacharya:2011ee,Donini:2012tt}.
We underline that while this work deals with charged current interactions, one can obtain valuable
information on active-sterile oscillations parameters also from the analysis of neutral current interactions (see~\cite{MINOS:2016viw,Adamson:2017uda,Adamson:2017zcg,Abe:2019fyx} for constraints
from existing data and~\cite{Coloma:2017ptb,Gandhi:2017vzo} for sensitivity studies of future experiments).

\section{Theoretical framework}

In the presence of a sterile neutrino, the flavor ($\nu_e,\nu_\mu,\nu_\tau, \nu_s$)
and the mass eigenstates ($\nu_1,\nu_2,\nu_3,\nu_4$) are mixed by a $4\times4$ 
unitary matrix
\begin{equation}
\label{eq:U}
U =   \tilde R_{34}  R_{24} \tilde R_{14} R_{23} \tilde R_{13} R_{12}\,, 
\end{equation} 
with $R_{ij}$ ($\tilde R_{ij}$) representing a real (complex) $4\times4$ rotation 
of a mixing angle $\theta_{ij}$ which endows the $2\times2$ submatrix 
\begin{eqnarray}
\label{eq:R_ij_2dim}
     R^{2\times2}_{ij} =
    \begin{pmatrix}
         c_{ij} &  s_{ij}  \\
         - s_{ij}  &  c_{ij}
    \end{pmatrix} ,
\,\,\,\,\,\,\,   
     \tilde R^{2\times2}_{ij} =
    \begin{pmatrix}
         c_{ij} &  \tilde s_{ij}  \\
         - \tilde s_{ij}^*  &  c_{ij}
    \end{pmatrix}
\,,    
\end{eqnarray}
in the  $(i,j)$ sub-block. For compactness, we have defined
\begin{eqnarray}
 c_{ij} \equiv \cos \theta_{ij}, \qquad s_{ij} \equiv \sin \theta_{ij}, \qquad  \tilde s_{ij} \equiv s_{ij} e^{-i\delta_{ij}}.
\end{eqnarray}
The matrix in Eq.~(\ref{eq:U}) posseses some useful properties: i) The 
3-flavor matrix is recovered if one sets $\theta_{14} = \theta_{24} = \theta_{34} =0$.
ii) For small values of the mixing angles $\theta_{14}$, $\theta_{24}$, and $\theta_{13}$, 
it is $|U_{e3}|^2 \simeq s^2_{13}$, $|U_{e4}|^2 = s^2_{14}$, 
$|U_{\mu4}|^2  \simeq s^2_{24}$, and $|U_{\tau4}|^2 \simeq s^2_{34}$, 
with a clear physical interpretation of the three mixing angles. 
iii) The leftmost positioning of the matrix $\tilde R_{34}$ guarantees
that the $\nu_{\mu} \to \nu_{e}$ probability
in vacuum is independent of $\theta_{34}$ and of the associated CP phase $\delta_{34}$
(see~\cite{Klop:2014ima}). 

For simplicity, we limit our treatment to the case of oscillations in vacuum.%
\footnote{In NO$\nu$A and T2K the matter effects induced by sterile neutrino oscillations  are very small.  
Notwithstanding, for the sake of precision, they are fully incorporated in our simulations.
A detailed description of the matter effects can be found in~\cite{Klop:2014ima}.}
As first shown in~\cite{Klop:2014ima}, the $\nu_{\mu} \to \nu_{e}$ 
probability can be expressed as the sum of three terms
\begin{eqnarray}
\label{eq:Pme_4nu_3_terms}
P^{4\nu}_{\mu e}  \simeq  P^{\rm{ATM}} + P^{\rm {INT}}_{\rm I}+   P^{\rm {INT}}_{\rm II}\,.
\end{eqnarray}
The first term, which is positive-definite, is related solely to the atmospheric mass-squared splitting and
gives the biggest contribution to the probability. It can be cast in the form
\begin{eqnarray}
\label{eq:Pme_atm}
 &\!\! \!\! \!\! \!\! \!\! \!\! \!\!  P^{\rm {ATM}} &\!\! \simeq\,  4 s_{23}^2 s^2_{13}  \sin^2{\Delta}\,,
 \end{eqnarray}
where $\Delta \equiv  \Delta m^2_{31}L/4E$ is the atmospheric oscillating factor, 
$L$  and $E$ being the neutrino baseline and energy, respectively. The other two terms
in Eq.~(\ref{eq:Pme_4nu_3_terms}) are generated by the interference of two frequencies 
and are not positive-definite. In particular, the second term in Eq.~(\ref{eq:Pme_4nu_3_terms})
arises from the interference of the solar and atmospheric frequencies and can be expressed as
\begin{eqnarray}
 \label{eq:Pme_int_1}
 &\!\! \!\! \!\! \!\! \!\! \!\! \!\! \!\! P^{\rm {INT}}_{\rm I} &\!\!  \simeq\,   8 s_{13} s_{12} c_{12} s_{23} c_{23} (\alpha \Delta)\sin \Delta \cos({\Delta + \delta_{13}})\,,
\end{eqnarray}
where we have  defined the ratio of the solar over the atmospheric mass-squared splitting, 
$\alpha \equiv \Delta m^2_{21}/ \Delta m^2_{31}$.
We recall that at the first  oscillation maximum it is $\Delta \sim \pi/2$.
The third term in Eq.~(\ref{eq:Pme_4nu_3_terms})
appears as a new genuine 4-flavor effect, and it arises from the interference of the
atmospheric frequency with the new frequency introduced by fourth
mass eigenstate. This term can be cast in the form~\cite{Klop:2014ima} 
\begin{eqnarray}
 \label{eq:Pme_int_2}
 &\!\! \!\! \!\! \!\! \!\! \!\! \!\! \!\! P^{\rm {INT}}_{\rm II} &\!\!  \simeq\,   4 s_{14} s_{24} s_{13} s_{23} \sin\Delta \sin (\Delta + \delta_{13} - \delta_{14})\,.
\end{eqnarray}
From inspection  Eqs.~(\ref{eq:Pme_atm})-(\ref{eq:Pme_int_2}), one can notice that the probability 
depends (besides the large mixing angle $\theta_{23}$) on three small mixing angles: 
$\theta_{13}$, $\theta_{14}$ and $\theta_{24}$.
One can notice that the values of such three mixing angles (estimated in the 3-flavor 
scenario~\cite{deSalas:2017kay,Capozzi:2018ubv,Esteban:2018azc} for $\theta_{13}$,
and in the 4-flavor framework~\cite{Capozzi:2016vac,Gariazzo:2017fdh,Dentler:2018sju,Diaz:2019fwt} 
for $\theta_{14}$ and $\theta_{24}$) are similar as one has 
$s_{13} \sim s_{14} \sim s_{24} \sim 0.15$. 
Therefore, one is legitimate to consider these three angles as small quantities
of the same order of magnitude $\epsilon$. We also note that  $|\alpha| \simeq \, 0.03$,
can be considered of order $\epsilon^2$. 
Hence, Eqs.~(\ref{eq:Pme_atm})-(\ref{eq:Pme_int_2}), show that the 
first (leading) contribution is of the second order, while each of the two interference terms 
is of the third order.

\section{Data used and details of the statistical analysis} 

We extracted the information relevant for NO$\nu$A and T2K from the data released 
in~\cite{NOVA_talk_nu2020} and~\cite{T2K_talk_nu2020}.
The analysis presented in this work is performed using the GLoBES 
software~\cite{Huber:2004ka,Huber:2007ji}, together with its new-physics package~\cite{Kopp:NSI}.
The sterile neutrino effects are included both in the $\nu_\mu \to \nu_e$ appearance channel, 
and in the $\nu_\mu \to \nu_\mu$ disappearance one. This holds both for neutrino and antineutrino modes. 
We have fixed the solar oscillation parameters $\theta_{12}$ and $\Delta m^2_{21}$ 
at their best fit values taken from the recent global fit~\cite{Capozzi:2020qhw}.
The constraint on the reactor mixing angle $\theta_{13}$ is incorporated as an external
gaussian prior likelihood estimated from~\cite{Capozzi:2020qhw}.
The atmospheric oscillations parameters $\theta_{23}$ and $\Delta m^2_{31}$ 
and the CP-phase $\delta_{13}$ are treated as free parameters and are marginalized away when needed. 
Concerning the mixing angles that involve the fourth state, 
we have fixed them to the values $\theta_{14} = \theta_{24}=8^\circ$ and $\theta_{34}=0$.
These are very close to the best estimates obtained in the global SBL analyses
performed within the 3+1 scheme~\cite{Capozzi:2016vac,Gariazzo:2017fdh,Dentler:2018sju,Diaz:2019fwt}.
Concerning the CP phase $\delta_{14}$, we vary its value in the range [$0,2\pi$].
The CP-phase $\delta_{34}$ has been taken equal to zero, but in any case
the analysis is not sensitive to it by construction, since we have set the
associated mixing angle $\theta_{34} = 0$.%
\footnote{We recall that in vacuum the $\nu_\mu \to \nu_e$ transition probability is independent of $\theta_{34}$
(and $\delta_{34}$). In matter, one expects a tiny dependence, which is very small in NO$\nu$A and in T2K
(see the appendix of~\cite{Klop:2014ima} for a detailed discussion). As shown in~\cite{Agarwalla:2016xxa}, at DUNE,
where matter effects are more pronounced, it will be possible to probe the CP-phase $\delta_{34}$ provided that $\theta_{34}$ is very large and close to its upper limit ($\sim 21^\circ$ at the 90\% C.L.~\cite{Dentler:2018sju}).}
We set the mass-squared splitting to $\Delta m^2_{41} = 1\,$eV$^2$, which
is close to the value currently indicated by the SBL data. However, we underline that our
findings would be unaltered for different values of such a parameter, provided 
that  $\Delta m^2_{41} > 0.1\,$eV$^2$. For such values, the fast
oscillations induced by the large mass-squared splitting get totally averaged
by the finite energy resolution of the detector. For the same motivation, 
the LBL setups are insensitive to the sign of $\Delta m^2_{41}$ and this allows us to
confine our study to positive values.  We have set the line-averaged constant Earth matter 
density%
\footnote{The line-averaged constant Earth matter density has been computed using the 
Preliminary Reference Earth Model (PREM)~\cite{PREM:1981}.} 
of 2.8 g/cm$^{3}$ for all the LBL experiments considered.
In our simulations, we have incorporated a full spectral analysis by making use of
the binned events spectra for each LBL experiment. In the statistical analysis, 
the Poissonian $\Delta\chi^{2}$ has been marginalized over the systematic 
uncertainties using the pull method as prescribed in Refs.~\cite{Huber:2002mx,Fogli:2002pt}.

\section{Numerical Results}

First of all, we comment on how much the quality of the fit improves in the 4-flavor case with respect to the
3-flavor scheme. We find a very minor improvement of $\Delta \chi^2 \sim 0.7$ in NO and  $\Delta \chi^2 \sim 0.9$ in IO. This means that the hypothesis of a light sterile neutrino is only weakly preferred over standard oscillations,
and signals that the 3+1 scenario, as we will discuss below in more detail, is not able to resolve the tension 
recently found between T2K and NO$\nu$A in the appearance channel.

\begin{figure}[t!]
\vspace*{-0.0cm}
\hspace*{-0.2cm}
\includegraphics[height=5.8cm,width=5.8cm]{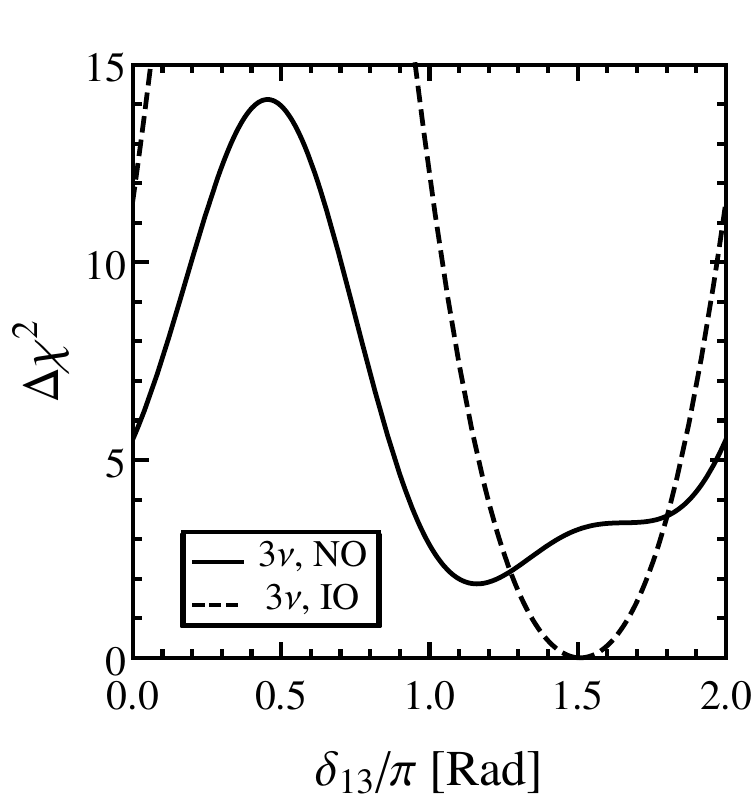}
\includegraphics[height=5.8cm,width=5.8cm]{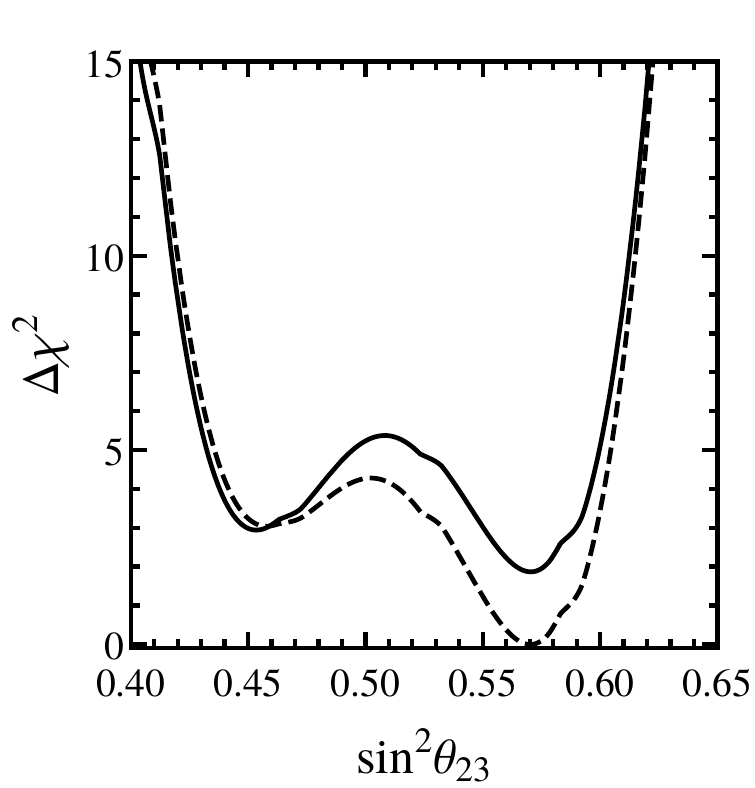}
\includegraphics[height=5.8cm,width=5.8cm]{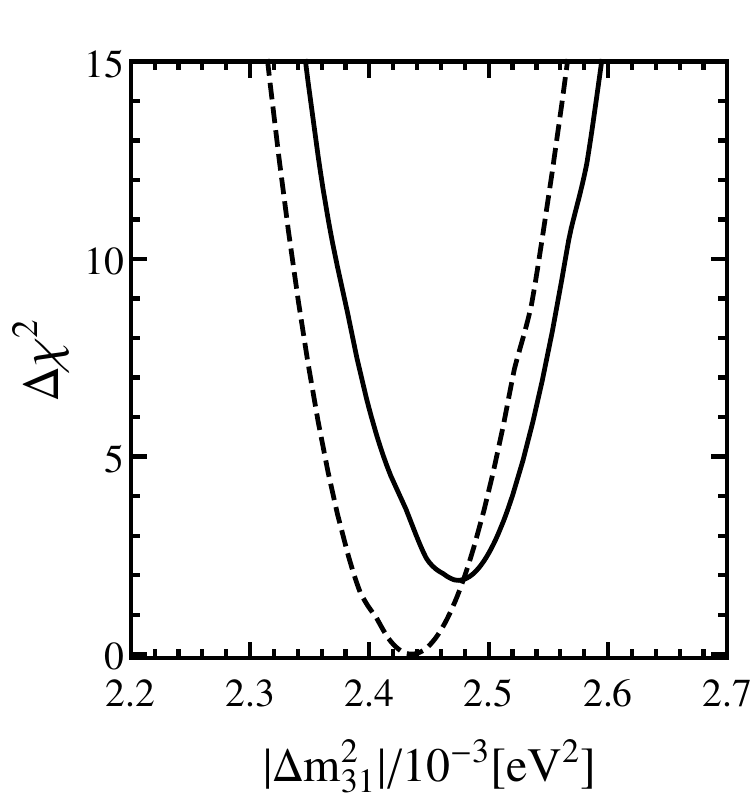}\\
\includegraphics[height=5.8cm,width=5.8cm]{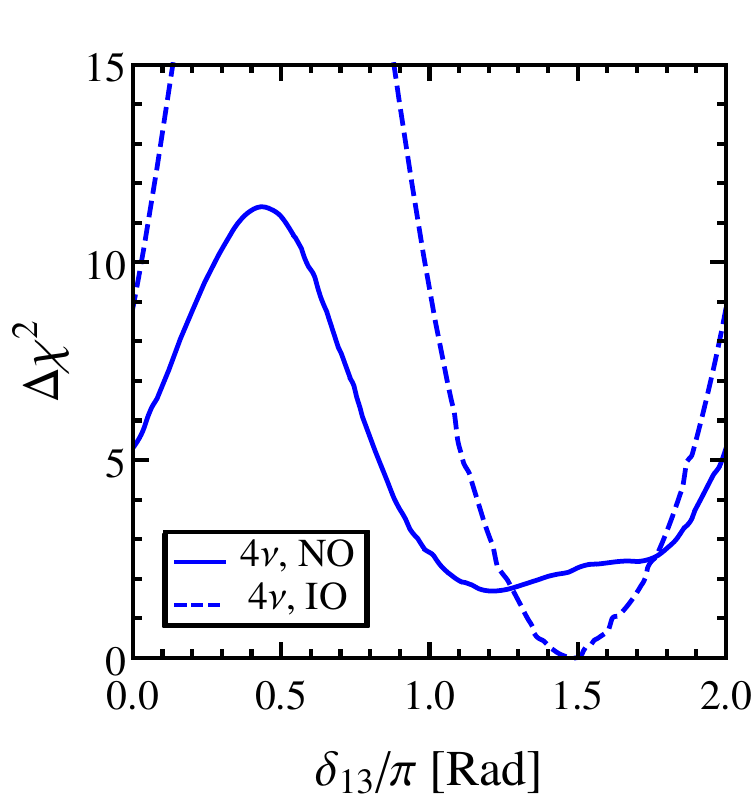}
\includegraphics[height=5.8cm,width=5.8cm]{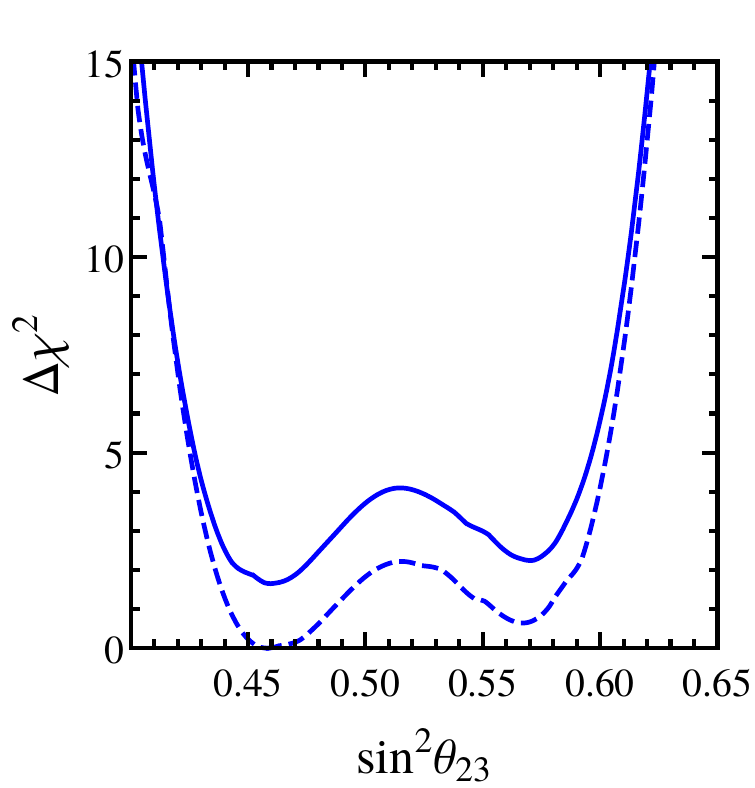}
\includegraphics[height=5.8cm,width=5.8cm]{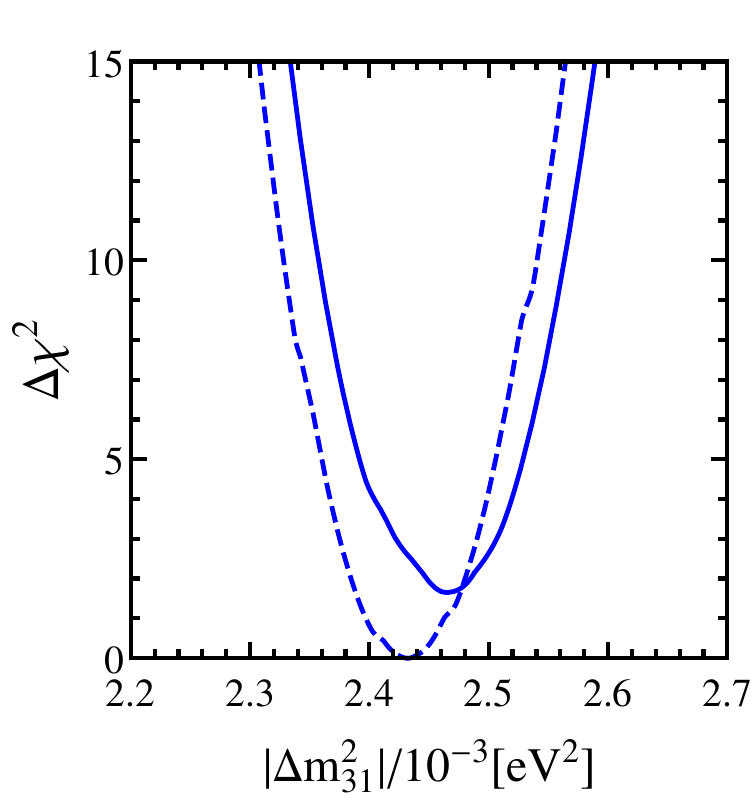}
\vspace*{-0.2cm}
\caption{Estimates of the oscillation parameters for the 3-flavor (upper panels) and 4-flavor 
(lower panels) scenarios determined by the combination of T2K and NO$\nu$A (with reactor constraint).
The continuous (dashed) curves refer to NO (IO).}
\label{fig:fit}
\end{figure}

The upper (lower) panels of Fig.~\ref{fig:fit} report the estimates of 
the oscillation parameters attained by expanding
the $\chi^2$ around the minimum value obtained when the 3-flavor (4-flavor) hypothesis is accepted as true.
The projections refer to $\delta_{13}$ (left panels),
$\theta_{23}$ (middle panels), and $|\Delta m^2_{31}|$ (right panels),
and are obtained by  combining T2K and NO$\nu$A data with an external prior on  $\theta_{13}$ 
as derived from reactors.
In all panels, the continuous (dashed) curve represents the NO (IO).
This figure shows that the 3-flavor ($\sim$1.4$\sigma$) weak indication in favor of the 
inverted ordering (also emerged in~\cite{Kelly:2020fkv}) remains almost unchanged 
in the 3+1 scheme ($\sim$1.3$\sigma$).
The inspection of the left panels shows that in the normal ordering, the preference in favor 
of $\delta_{13} \sim 1.2 \pi$ found in the 3-flavor framework is basically unaltered 
 in the 3+1 scheme. We stress that such a value of $\delta_{13}$ is the result of a compromise
 between T2K, which prefers $\delta_{13} \sim 1.5 \pi$ and NO$\nu$A, which prefers 
 $\delta_{13} \sim 0.8 \pi$. The fact that the best fit  value in the 3+1 scheme does not change,
 signals that light sterile neutrinos are not able to resolve the tension among T2K and NO$\nu$A.
 This is analogous to what has been recently noticed for the scenario of non-unitary neutrino mixing~\cite{Forero:2021azc}.
 Therefore, at the moment, the only mechanism able to explain the tension is given 
 by hypothetical neutrino non-standard interactions (NSIs), as recently shown in~\cite{Chatterjee:2020kkm} (see also~\cite{Denton:2020uda}).
 This different capability of NSIs to resolve the discrepancy between the two experiments stems from their
 dependence on matter effects, to which T2K and NO$\nu$A have a different sensitivity, as already
 noticed in a previous work~\cite{Capozzi:2019iqn}. In contrast,
 in the case of  sterile neutrinos and non-unitary neutrino mixing only kinematical (vacuum) effects are involved.

Concerning the atmospheric mixing angle $\theta_{23}$ (middle panels) we see that, regardless
of the NMO, the best fit lies in the higher octant ($\sin^2\theta_{23}\sim 0.57$) in the 3-flavor case.
In the 4-flavor scenario, the preference swaps to the lower octant ($\sin^2\theta_{23}\sim 0.46$).
This is in agreement with the behavior predicted in the forecast study~\cite{Agarwalla:2016xlg},
which evidenced that sterile neutrinos may hamper the determination of the octant of $\theta_{23}$
even in future LBL experiments. Finally, in the right panels we report the estimate of the atmospheric 
mass-squared $\Delta m^2_{31}$ splitting. From these plots we can observe that both in NO and IO
 the best fit values and allowed ranges are very similar in the 3-flavor and 4-flavor schemes.

\begin{figure}[t!]
\vspace*{0.0cm}
\hspace*{-0.1cm}
\includegraphics[width=6.5 cm]{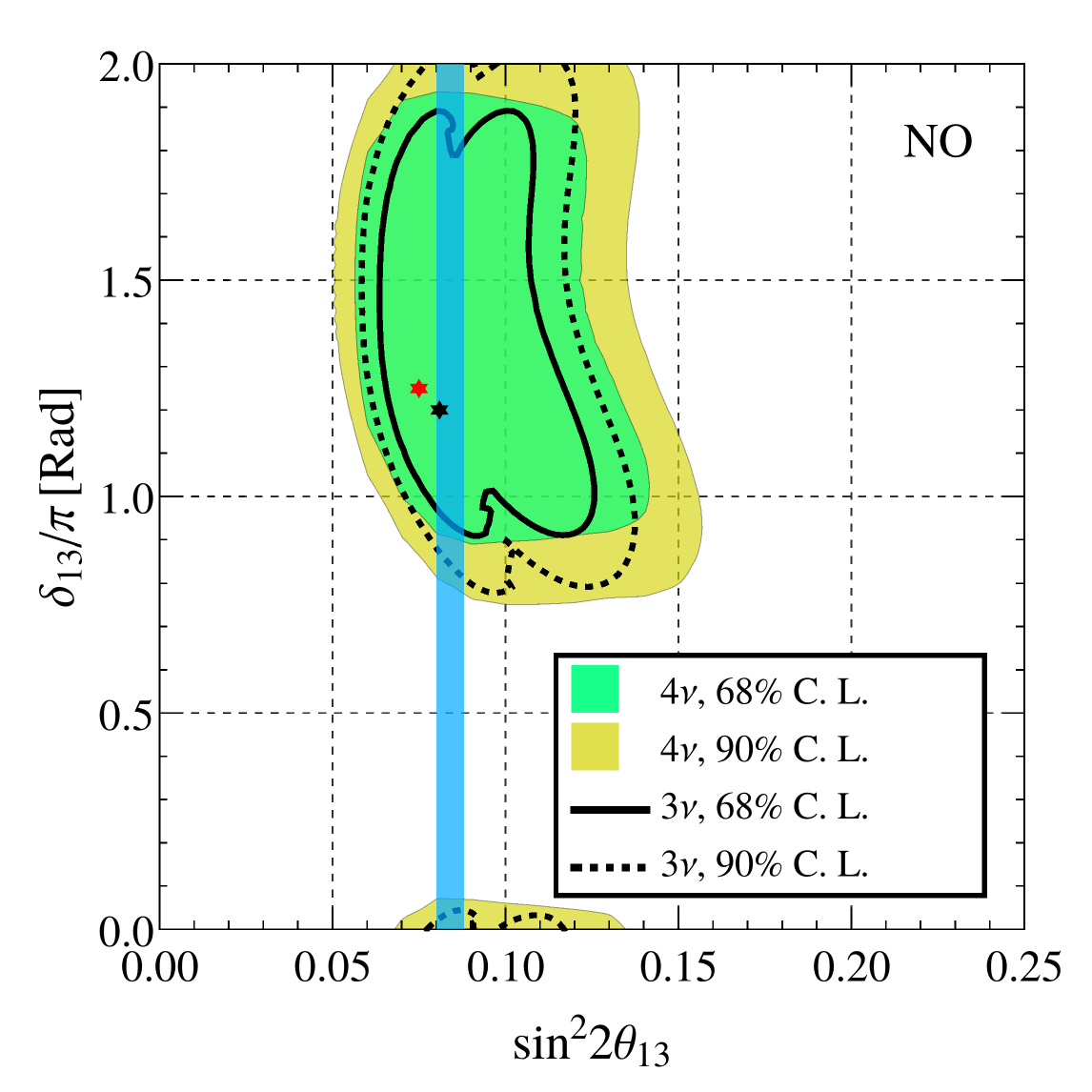}
\includegraphics[width=6.5 cm]{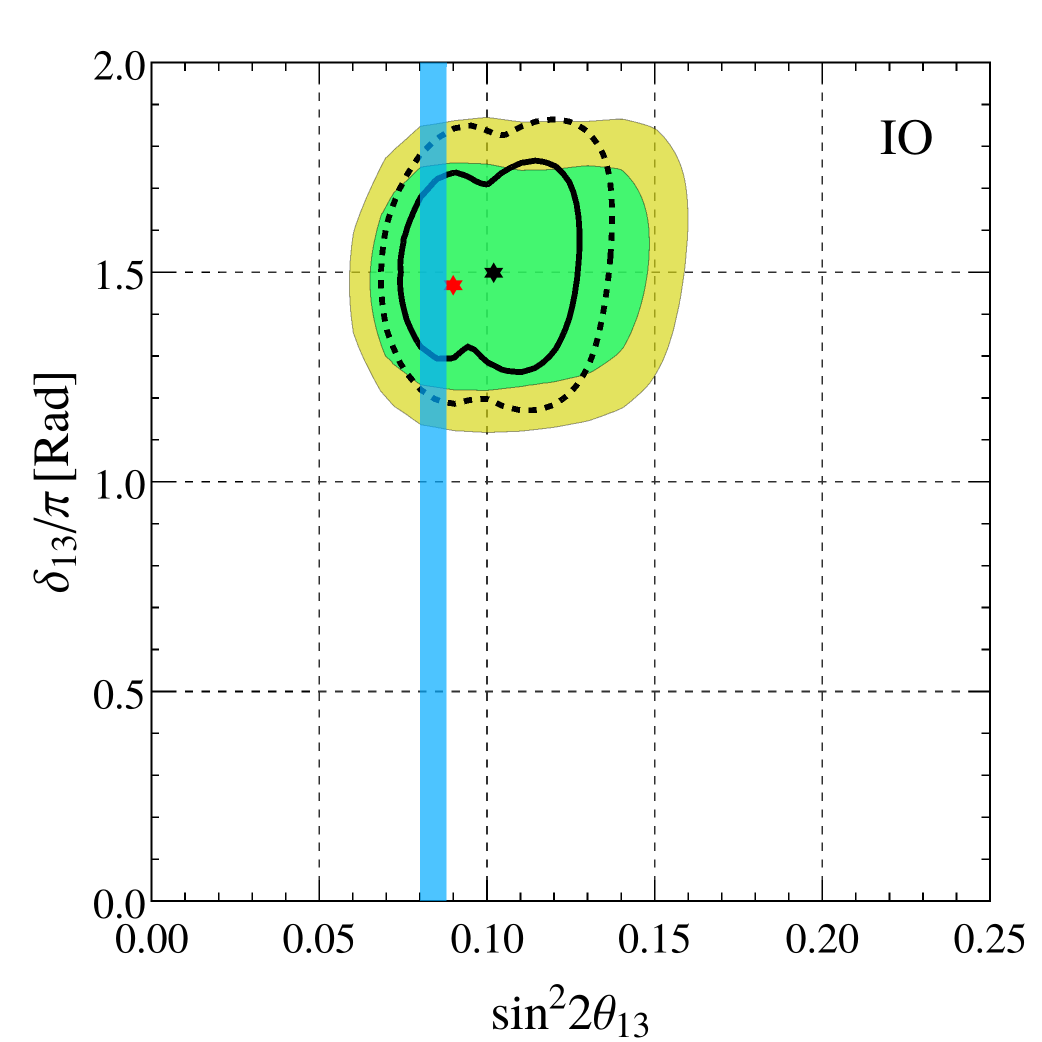}
\vspace*{0.0cm}
\caption{Allowed regions obtained  from the combination of T2K and NO$\nu$A  (without reactor constraint)
in the plane spanned by $\sin^2 2 \theta_{13}$ and $\delta_{13}$. Left (right) panel refers to NO (IO).
The black contours represent the 3-flavor case, while the filled regions pertain to the 4-flavor scheme.
The black (red) star indicates the best fit point in the 3-flavor (4-flavor) scenario.
The vertical band indicates the 1$\sigma$ constraint on  $\theta_{13}$ determined by reactor experiments. 
\label{fig:corr_13_d13}}
\end{figure}

In order to better understand how the 1-dimensional constraints shown in Fig.~\ref{fig:fit} arise,
it is useful to look at the 2-dimensional pojections in the plane spanned by a couple of
oscillation parameters. In addition, such plots will enrich our comprehension by evidencing
potential correlations among the oscillation parameters. Figure~\ref{fig:corr_13_d13} displays the constraints 
in the plane spanned by $\theta_{13}$  and $\delta_{13}$. Left (right) panel refers to NO (IO). 
The black contours represent the 3-flavor case, while the filled regions refer to the 4-flavor scenario. 
In both cases we show the 68\% and 90\% C.L. for two d.o.f.. Differently from all other plots,
 in this figure, we have drawn the regions allowed by NO$\nu$A and T2K {\em without} 
 the external prior on $\theta_{13}$ coming from reactor experiments. Such a prior
 is shown at the 1$\sigma$ level as a thin vertical band in each of the two panels
 of Fig.~\ref{fig:corr_13_d13}. This figure allows us to appreciate possibles synergies between
 reactor and accelerator constraints. In particular, one can observe that the level of superposition
 of the reactor band with the allowed regions of the LBL experiments, is comparable in the NO
 and IO cases. Therefore, the inclusion of the reactor prior on $\theta_{13}$ has no role
 in the preference of IO over NO observed in the 1-dimensional projections shown in Fig.~\ref{fig:fit},
 which entirely comes from the LBL experiments T2K and NO$\nu$A. In the 4-flavor case the
 agreement between reactors and LBL experiments is basically unaltered, so one
 expects that IO will be preferred over NO also in this enlarged framework, as found in the
 1-dimensional projections shown in Fig.~\ref{fig:fit}. Concerning the CP-phase $\delta_{13}$,
 we see that the two LBL experiments  tend to favor values of $\delta_{13}$ close
 to $\sim1.2\pi$ for NO and $\sim1.5\pi$ in IO. The combination with reactors has almost
 no role also on this parameter, which currently is entirely determined by the LBL data.
  
\begin{figure}[b!]
\vspace*{0.0cm}
\hspace*{-0.1cm}
\includegraphics[width=6.5 cm]{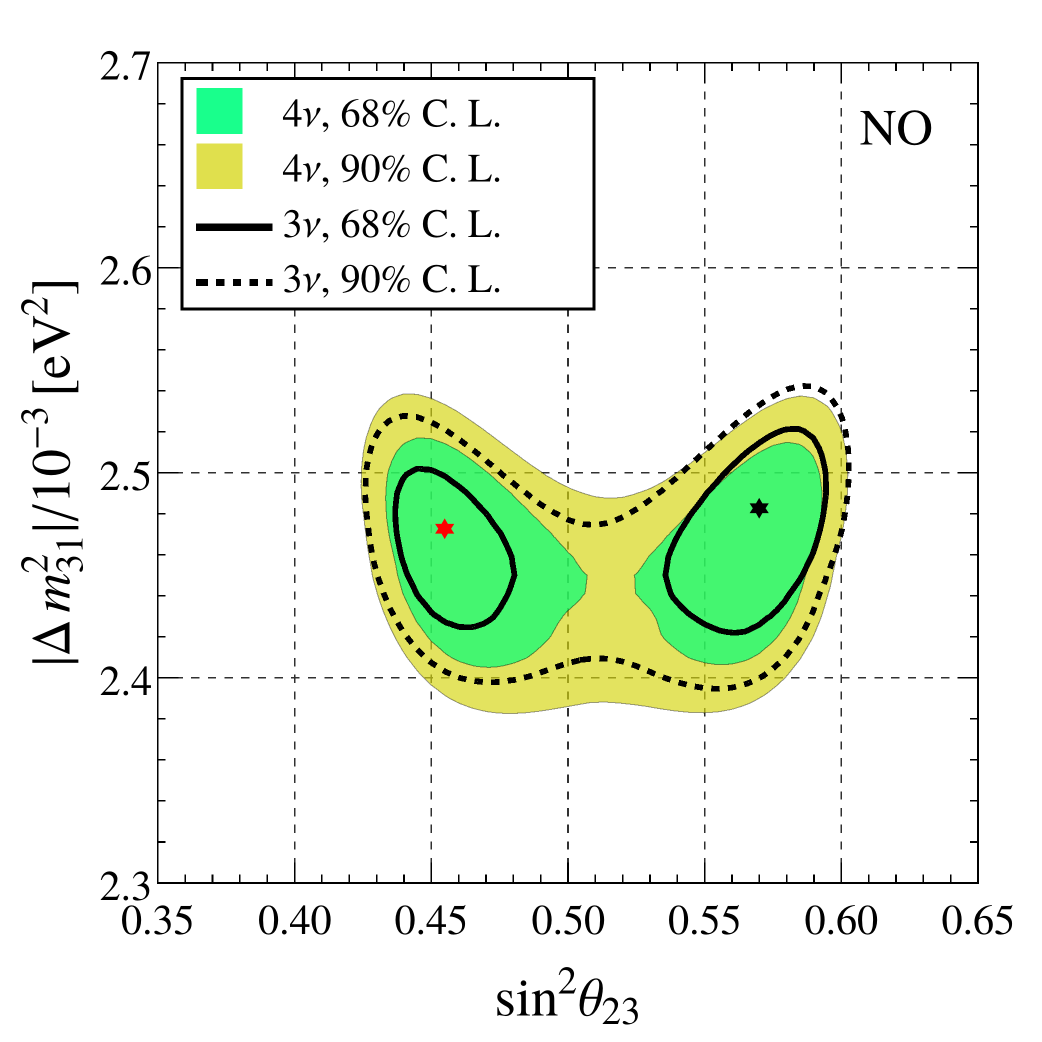}
\includegraphics[width=6.5 cm]{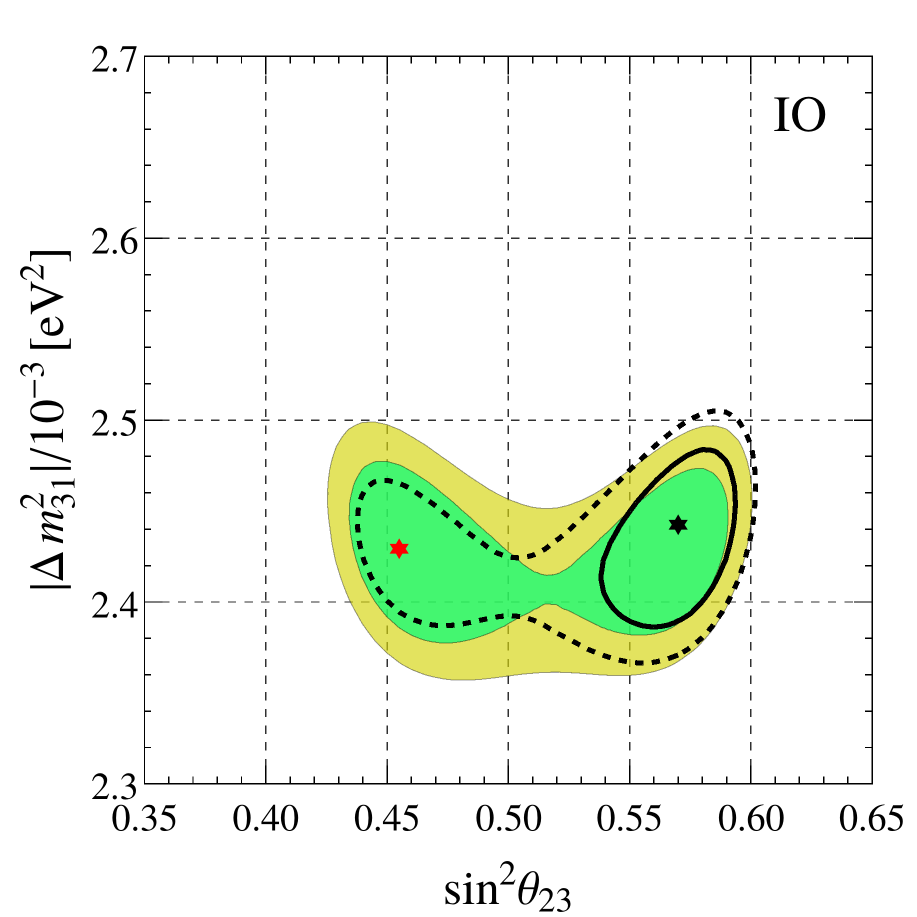}
\vspace*{0.0cm}
\caption{Allowed regions obtained  from the combination of T2K and NO$\nu$A (with reactor constraint)
in the plane spanned by $\sin^2\theta_{23}$ and $|\Delta m^2_{31}|$. Left (right) panel refers to NO (IO).
The black contours represent the 3-flavor case, while the filled regions pertain to the 4-flavor scheme.
The black (red) star indicates the best fit point in the 3-flavor (4-flavor) scenario.
\label{fig:corr_23_D31}}
\end{figure}  

\begin{figure}[t!]
\vspace*{0.0cm}
\hspace*{-0.1cm}
\includegraphics[width=6.5 cm]{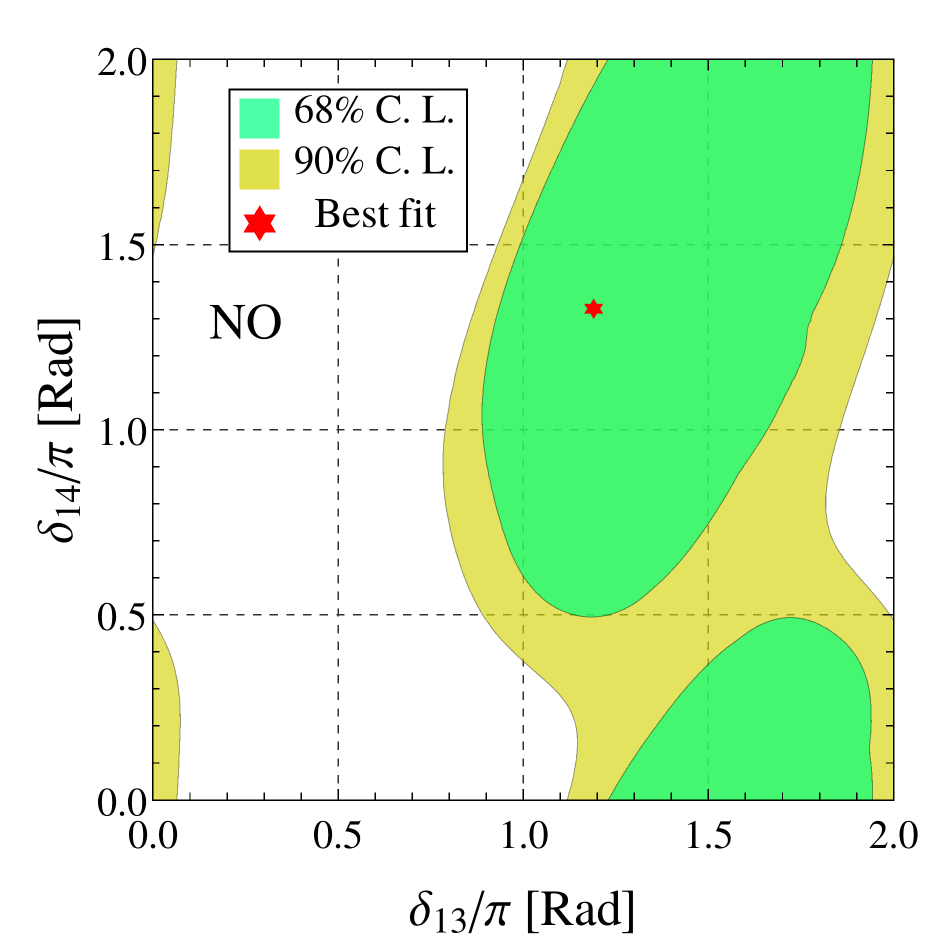}
\includegraphics[width=6.5 cm]{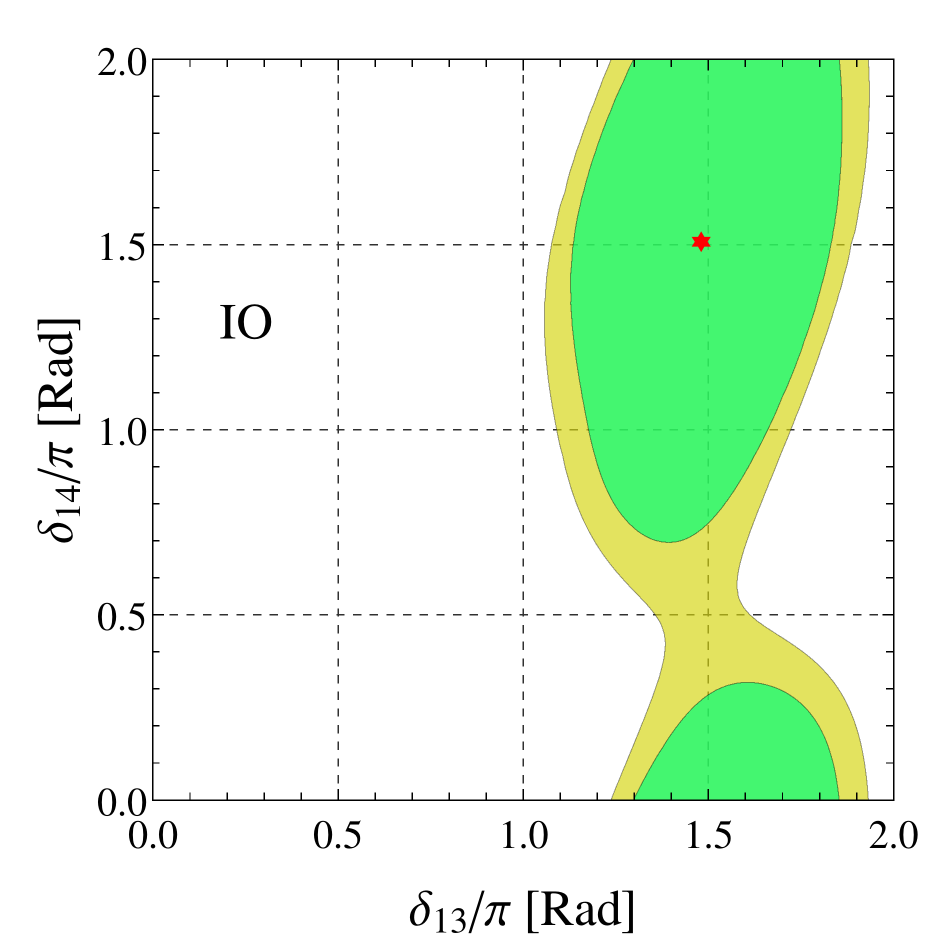}
\vspace*{0.0cm}
\caption{Regions allowed by the combination of T2K and NO$\nu$A  (with reactor constraint)
 in the plane spanned by the 
two CP-phases $\delta_{13}$ and $\delta_{14}$. Left (right) panel refers to NO (IO).
\label{fig:CPcorr}}
\end{figure}

Figure~\ref{fig:corr_23_D31} shows the constraints in the plane spanned by $\theta_{23}$  and $|\Delta m^2_{31}|$.
Left (right) panel refers to NO (IO). The black contours represent the 3-flavor case, while
the filled regions refer to the 4-flavor scenario. In both cases we show the 68\% and 90\% C.L. for two d.o.f..
As expected, the allowed regions are larger in  4-flavor than in 3-flavor scheme.
This is natural as one has more freedom in the fit
in the 3+1 scheme. However, one can notice that the range allowed for $|\Delta m^2_{31}|$
is very similar in the two scenarios both in NO and IO. This feature can be understood recalling that the  
$\Delta m^2_{31}$ is constrained mostly by the disappearance channel, which probes
the $\nu_\mu \to \nu_\mu$ survival probability. This in turn is almost insensitive to the
presence of sterile neutrinos. 
Figure~\ref{fig:CPcorr} displays the constraints in the plane spanned by the two CP-phases 
$\delta_{13}$ and $\delta_{14}$ for NO (left panel) and IO (right panel). We can observe
that although there is an appreciable sensitivity to the new CP-phase   $\delta_{14}$,
it is less constrained with respect to the 3-flavor CP-phase $\delta_{13}$.
We note that in NO the CP-conserving case ($\delta_{13}, \delta_{14})=(\pi, \pi)$ is allowed at the $68\%$
confidence level. In contrast, in the IO case there is a appreciable rejection of both  
the CP-conserving cases ($\delta_{13}, \delta_{14})=(0,0)$ and $\delta_{14} = (\pi, \pi)$.
We must observe that although the two experiments T2K and NO$\nu$A prefer the IO, 
all the other data do not support this indication. In fact, all the recent global analyses~\cite{Esteban:2020cvm,deSalas:2020pgw,marrone_talk_neutel_2021} (which include the latest data of T2K and NO$\nu$A) evidence a 
preference for NO at the $\sim2.7\sigma$  level. If such a global preference in favor of NO will be confirmed
by future data, it would mean that the preferred regions for the CP-phases would be those reported in the left panel of Figure~\ref{fig:CPcorr}.
In this case, we would have no hint in favor of CPV,  for both $\delta_{13}$ and $\delta_{14}$.

\section{Conclusions}

We have considered the impact of a light sterile neutrino in the interpretation of the latest data 
of the long baseline experiments NO$\nu$A and T2K. We have assessed the estimates of the 
standard 3-flavor parameters, highlighting the perturbations induced in the 4-flavor scheme. 
We find that all the basic features of the 3-flavor analysis, including the weak 
indication ($\sim$1.4$\sigma$) in favor of the inverted neutrino mass ordering, the preference for values of the CP-phase $\delta_{13} \sim 1.2\pi$, and the substantial degeneracy of the two octants of $\theta_{23}$, all remain almost unchanged in the 4-flavor scheme. Our work also demonstrates that it is possible to attain some constraints on the new CP-phase $\delta_{14}$. Finally, we have pointed out that, differently from non-standard neutrino interactions, light sterile neutrinos are not capable to alleviate the tension recently emerged between NO$\nu$A and T2K  in the appearance channel. We hope that our work may trigger more complete 4-flavor analyses incorporating also the atmospheric neutrino data, which are expected to be sensitive both to the NMO and to the standard and non-standard CP-phases $\delta_{13}$ and $\delta_{14}$. However, we point out, that an accurate 4-flavor analysis of the current atmospheric data can be performed only from inside the experimental
collaborations. To this regard, we would like to underline that the publication by the experimental collaborations 
of a $\chi^2$ map for the 4-flavor analysis (as already available for the 3-flavor case), would be  extremely useful, since it would allow one to perform a global analysis of all data sensitive to the NMO and (ordinary and new) CP-phases. In the eventuality of a discovery of a light sterile neutrino at SBL experiments, the availability
of such pieces of information would become an indispensable tool for exploring the 4-flavor framework.  

\section*{Acknowledgments}
A.P. acknowledges partial support by the research grant number 
2017W4HA7S ``NAT-NET: Neutrino and Astroparticle Theory Network'' 
under the program PRIN 2017 funded by the 
Italian Ministero dell'Istruzione, dell'Universit\`a e della Ricerca (MIUR) 
and by the research project {\em TAsP} funded 
by the Instituto Nazionale di Fisica Nucleare (INFN).

\bibliographystyle{JHEP_new}
\bibliography{Sterile-References_2020}

\end{document}